\pdfoutput=1 
\documentclass[twocolumn,superscriptaddress]{revtex4-1}
\usepackage{amsmath,amsfonts}
\usepackage{enumerate}
\usepackage[utf8]{inputenc}
\usepackage{xcolor}
\usepackage{hyperref}
\hypersetup{colorlinks,breaklinks,allcolors=blue!60!black}
\usepackage{graphicx}

\begin{document}
\title{Bell violation with entangled photons, free of the
  coincidence-time loophole}

\author{Jan-\AA{}ke Larsson} \email{jan-ake.larsson@liu.se}
\affiliation{Institutionen för Systemteknik, Linköpings Universitet,
  581 83 Linköping, Sweden}

\author{Marissa Giustina}%
\affiliation{Institute for Quantum Optics and Quantum Information
  (IQOQI), Austrian Academy of Sciences, Boltzmanngasse 3, 1090
  Vienna, Austria}%

\author{Johannes Kofler}%
\affiliation{Max Planck Institute of Quantum Optics (MPQ),
  Hans-Kopfermannstra{\ss}e 1, 85748 Garching/Munich, Germany}%

\author{Bernhard Wittmann}%
\affiliation{Institute for Quantum Optics and Quantum Information
  (IQOQI), Austrian Academy of Sciences, Boltzmanngasse 3, 1090
  Vienna, Austria}%

\author{Rupert Ursin}%
\affiliation{Institute for Quantum Optics and Quantum Information
  (IQOQI), Austrian Academy of Sciences, Boltzmanngasse 3, 1090
  Vienna, Austria}

\author{Sven Ramelow} \email{sven.ramelow@univie.ac.at}
\affiliation{Institute for Quantum Optics and Quantum Information
  (IQOQI), Austrian Academy of Sciences, Boltzmanngasse 3, 1090
  Vienna, Austria}%
\affiliation{Vienna Center for Quantum Science and Technology (VCQ),
  Faculty of Physics, University of Vienna, Boltzmanngasse 5, A-1090
  Vienna, Austria}%
\affiliation{Cornell University, 271 Clark Hall, 142 Science Dr.,
  Ithaca, 14853 NY, USA}%

\date{\today}
\begin{abstract}
  In a local realist world view, physical properties are defined prior
  to and independent of measurement, and no physical influence can
  propagate faster than the speed of light. Proper experimental
  violation of a Bell inequality would show that the world cannot be
  described within local realism. Such experiments usually require
  additional assumptions that make them vulnerable to a number of
  ``loopholes.'' A recent experiment \cite{Gius2013} violated a Bell
  inequality without being vulnerable to the detection (or
  fair-sampling) loophole, therefore not needing the fair-sampling
  assumption. Here we analyze the more subtle coincidence-time
  loophole, and propose and prove the validity of two different
  methods of data analysis that avoid it. Both methods are general and
  can be used both for pulsed and continuous-wave experiments. We
  apply them to demonstrate that the experiment mentioned above
  violates local realism without being vulnerable to the
  coincidence-time loophole, therefore not needing the corresponding
  fair-coincidence assumption.
\end{abstract}
\maketitle

When attempting to provide a conclusive answer to the question by
Einstein, Podolsky, and Rosen (EPR), ``Can [the] quantum-mechanical
description of physical reality be considered complete?''
\cite{Eins1935}, it is important that assumptions concerning the
physical reality are kept to an absolute minimum.

The usual assumptions underlying Bell inequalities are those of local
realism --- namely that properties of physical systems are elements of
reality, and that these cannot be influenced faster than the speed of
light; and freedom of choice --- namely that the measurement setting
choices are independent of the hidden variables and vice versa
\cite{Bell1964,Bell1990}. However, all experimental demonstrations
that attempt to violate a Bell inequality to date have needed
additional assumptions in order to claim the invalidity of local
realism.  In principle, such a violation could be caused by the
failure of these additional assumptions, rather than the more
fundamental assumptions of local realism.

In experiments involving photons, a well known problem is that not all
photons emitted by the source actually are registered in the
detectors. The problem is usually referred to as the ``fair-sampling''
or ``detection'' or ``detector-efficiency'' loophole but really
concerns the efficiency of the entire experimental setup, since there
can be various causes for missing detections. The outcomes that are
registered might display correlations that violate the Bell inequality
even though the experiment can be described by a local realist model
\cite{Pear1970}.  The ``fair-sampling'' assumption is often used in
this situation, often motivated by the assumption that successful
photon detection depends only on the hidden variable and not the
measurement setting. Fair sampling means that the observed outcomes of
detected photons faithfully reproduce the outcome statistics of all
emitted photons, if they all had been detected. This assumption is not
needed in high-efficiency experiments using, e.g., atoms or
superconducting qubits \cite{Rowe2001,Ansm2009} and the
detector-efficiency loophole has only recently been avoided in
photonic violations of the Clauser-Horne (CH) inequality
\cite{Clau1974} (or Eberhard inequality \cite{Eber1993}), which do not
require the fair-sampling assumption \cite{Gius2013,Chri2013}.

Other common assumptions include ``locality'' \cite{Aspe1982,Weih1998}
and ``freedom of choice'' \cite{Sche2010}; assumptions may also refer
to properties of decaying systems \cite{Hies2012} or properties of
photons \cite{Fran1989,Aert1999}, and the list continues. Any of these
makes an experiment vulnerable to explanation by a local realist
model. Avoiding all these assumptions simultaneously in one experiment
--- usually called a ``loophole-free'' or ``definitive'' Bell test ---
remains an open task.

One less-known but equally serious problem is that of identifying
which outcomes belong together \cite{Lars2004}, sometimes referred to
as the ``coincidence-time'' loophole. Both the EPR paradox and the
Bell inequality concern \emph{pairs of outcomes} at two remote
sites. In an experiment, it is therefore necessary to identify which
outcomes make up a pair, which may be a non-trivial task. Commonly, in
photon experiments, relative timing is used to identify pairs: if two
clicks are close in time they are ``coincident,'' otherwise they are
not. The problem of pair identification is especially pronounced in
continuously pumped photonic experiments, but is in principle present
in all experiments that have rapid repetition in the same physical
system.

In this situation, it may happen that some pairs are not identified
correctly. For example, if the detector jitter is large and the
coincidence window small, some pairs will not register as
coincident. Loss of coincidences would reduce the subset of registered
pairs, and the remaining coincidences might display correlations that
enable a Bell violation even though the experiment can be described by
a local realist model \cite{Lars2004}. When coincidences are lost, a
``fair-coincidence'' assumption is needed: that the observed outcomes
of all \emph{identified pairs} faithfully reproduce the outcome
statistics of all \emph{detected pairs of photons}, if they all had
been \emph{correctly identified}.  This can be motivated by the
assumption that the time of an individual photon detection depends
only on the hidden variable, and not the measurement setting.

% The assumption of fair coincidences is a much stronger assumption
% than that of fair sampling, which only refers to detection or
% non-detection.
Historically, only the fair-sampling assumption has been explicitly
made, and identification of pairs within the available measurement
data has not been thought of as a problem. In fact, until 2003-2004
\cite{Lars2004}, it was thought to be trivially true that coincidence
determination, or timing, had no detrimental effect on Bell tests of
local realism whatsoever. But since this is not the case, the
fair-coincidence assumption must have been tacitly made in every
experiment to date, with only a few recent exceptions
\cite{Ague2012,Chri2013}.  Here, we will provide a proof that the
continuously pumped experiment \cite{Gius2013} also does not need the
fair-coincidence assumption, since it is actually not vulnerable to
the loophole.

The remainder of this note will I) be a more complete description of
the coincidence-time loophole and how to avoid it, followed by II) an
analysis of the data from \cite{Gius2013} showing that there indeed is
a violation of local realism, free of the fair-coincidence assumption.

\begin{figure}[t]
  \centering
  \hypertarget{fig:1}{}%
  \includegraphics[width=\linewidth]{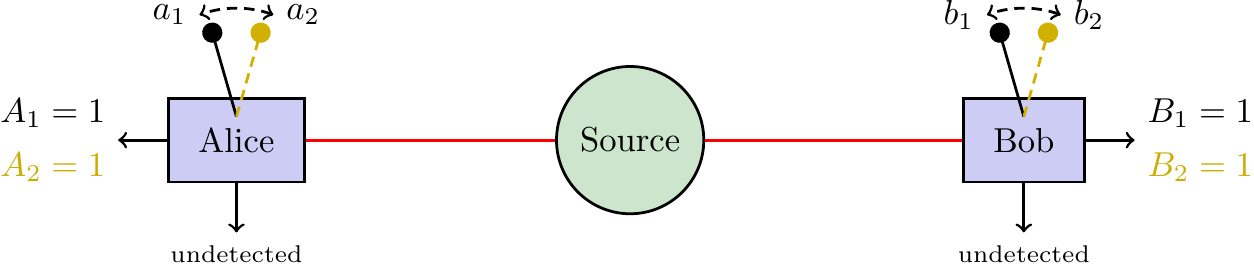}
  \caption{Principle of the experiment. Violation of the Clauser-Horne
    inequality needs an EPR (Einstein, Podolsky, Rosen) source of
    entangled pairs and two-setting measurement devices. Here,
    polarization entanglement and measurement is used. Each
    measurement device can rotate the photon's polarization according
    to one of two settings ($a_1$ or $a_2$ and $b_1$ or $b_2$) before checking
    if a photon arrives at the ``ordinary'' output of a polarizing
    beam splitter. This is recorded as the event $A_1=1$, $A_2=1$, $B_1=1$
    or $B_2=1$, depending on site and setting, as appropriate.}
  \label{fig:1}
\end{figure}

\begin{figure}[t]
  \centering
  \hypertarget{fig:2}{}%
  \includegraphics[width=\linewidth]{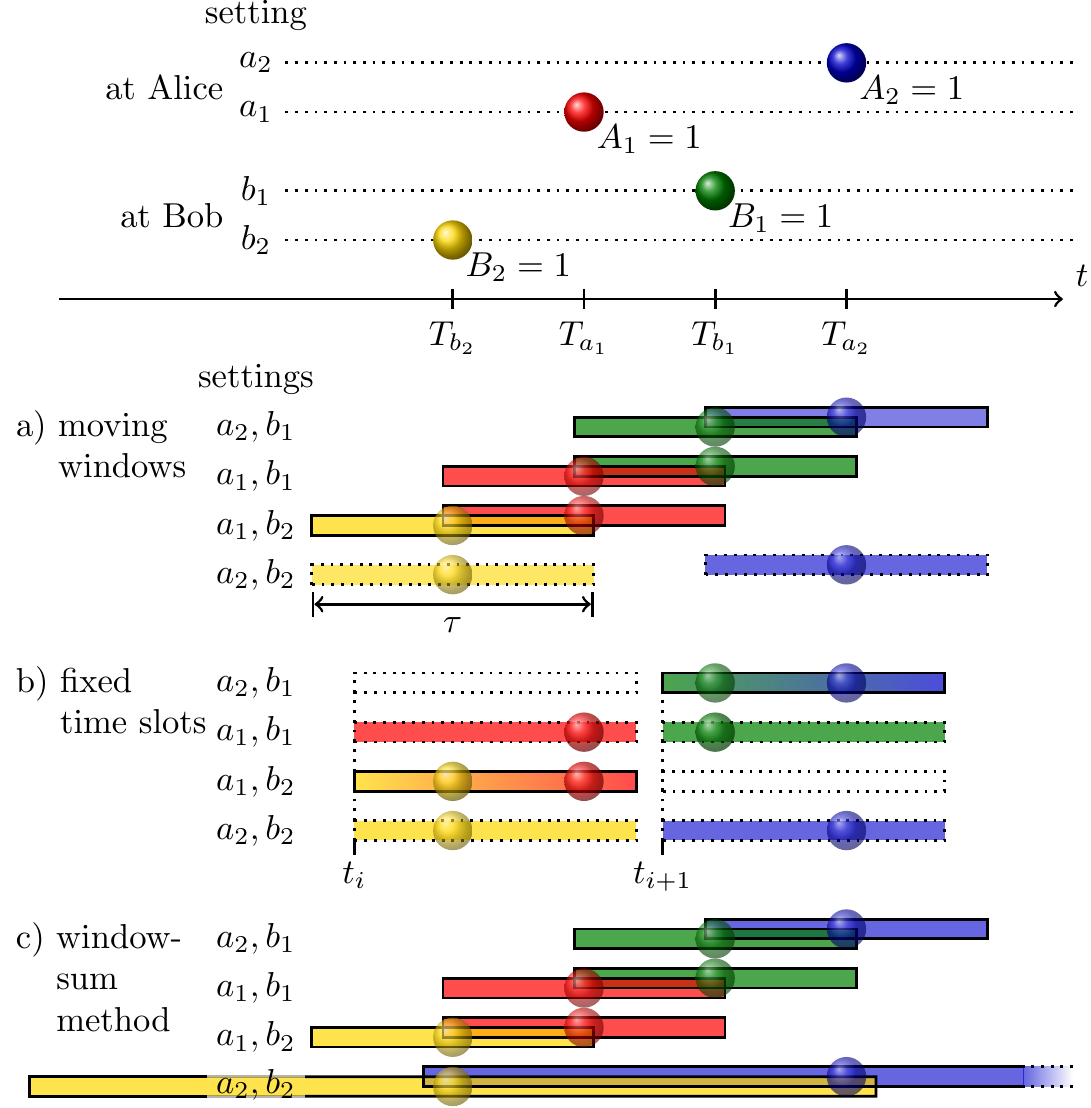}
  \caption{Comparison of the three different methods to identify
    coincidences. At the top is an example of a possible assignment to
    the detection times for a given $\lambda$, chosen to illustrate
    the key features of the different pair identification methods. In
    method \textbf{a)}, a coincidence is identified if detection times
    are close enough (with difference of at most half the coincidence
    window $\tau$), here for settings $a_2,b_1$; $a_1,b_1$; and
    $a_1,b_2$ but not for $a_2,b_2$. The $a_2,b_2$ events would then
    be misidentified as not being coincident. The fair-coincidence
    assumption implies that this, on average, happens equally often
    for all combinations, which is a substantial restriction on
    possible local realist models. Method \textbf{b)} uses fixed time
    slots of size $\tau$ to avoid the coincidence-time loophole by
    ensuring that if $a_2,b_2$ would not give a coincidence, at least
    one of $a_2,b_1$; $a_1,b_1$; or $a_1,b_2$ would also not do so (in
    the figure, this is $a_1,b_1$). Above, the time slots are
    separated as is appropriate for a pulsed experiment, while
    adjacent slots is the best choice for a continuously pumped
    experiment. Method \textbf{c)} uses a different approach, namely a
    longer time window for the $a_2,b_2$ settings, to ensure that if
    \emph{all} the others ($a_2,b_1$; $a_1,b_1$; or $a_1,b_2$) would
    give a coincidence, so would $a_2,b_2$.  The two new methods are
    opposite in the sense that in method b) coincidences are lost as
    compared to method a) while in method c) coincidences are gained;
    both avoid the coincidence-time loophole. }
  \label{fig:2}
\end{figure}

\textbf{Avoiding the coincidence-time loophole ---} In a typical
Bell-type experiment based on photon pairs (see
\hyperlink{fig:1}{Fig.~\ref*{fig:1}}), the usual way to determine if
two events belong to the same pair is to compare the two detection
times and conclude that a coincidence has happened if the two times
are close. More precisely, there is a coincidence for setting $a
\in \{a_1,a_2\}$ at the first site (Alice) and $b \in \{b_1,b_2\}$
at the second (Bob) if their detection times $T_A$ and $T_B$ are
close enough so that
\begin{equation}
  \label{eq:window}
  \big|T_{A}(a)-T_{B}(b)\big| < \tfrac{1}{2} \tau.
\end{equation}
Here, the chosen coincidence window width $\tau$ is the total possible
deviation in a detection time at one site, given the detection time at
the other.  To the experimentalist, this is the event ``there is a
coincidence,'' and the probability of such an event is well-defined
even without reference to hidden variables (see
\hyperlink{fig:2}{Fig.~\ref{fig:2}a}).

In a local realist model, the detection times $T_A$ and $T_B$ at the
two sites would be random variables that depend on the local settings
$a$ and $b$, with the hidden variable $\lambda$ as argument:
$T_{A}(a,\lambda)$, and $T_{B}(b,\lambda)$. Here, the locality
assumption ensures that the detection times do not depend on the
remote setting, just as is assumed for the outcomes.  Coincidences (of
outcomes $A \in \{A_1,A_2\}$ for Alice and $B \in \{B_1,B_2\}$ for
Bob, where the subscript denotes the setting used for measurement) now
occur on a subset of the hidden-variable space that can be written as
\begin{equation}
  \label{eq:6}
  \Lambda_{A_jB_k}
  =\big\{\lambda:\big|T_{A}(a_j,\lambda)-T_{B}(b_k,\lambda)\big|
  <\tfrac{1}{2} \tau\big\},
\end{equation}
with $j,k = 1,2$. This is the mathematical, or rather, the
probabilistic formalization of the coincidence event, and the
mathematical term for such a subset of the sample space is
``event.''

The structure of the above coincidence set is very different from
the structure of the set of coincidences when only missing outcomes
(and possibly unfair sampling) are taken into account. If clicks
occur on the sets $\Lambda_{A_j}$ and $\Lambda_{B_k}$, and all
coincidences are correctly identified, the coincidence set has the
factorizable structure
\begin{equation}
  \Lambda_{A_jB_k}=\Lambda_{A_j}\cap\Lambda_{B_k}.\label{eq:2}
\end{equation}
This leads to an experimental efficiency requirement of at least
$\approx$~$82.84\%$ to achieve a violation of local realism (in the
Clauser-Horne-Shimony-Holt \cite{Clau1969} inequality) free of the
fair-sampling assumption. Because the set \eqref{eq:6} cannot be
factored, the bound to avoid the fair-coincidence assumption will be
higher, $\approx$~$87.87\%$ \cite{Lars2004}.

In this note, we are interested in the Clauser-Horne (CH) inequality,
which holds under the assumptions of \emph{realism}, \emph{locality},
and \emph{freedom of choice};
\begin{align}
    P&(A_1=B_2=1) + P(A_2=B_1=1)-P(A_2=B_2=1)\notag\\
    &\le P(A_1=1) + P(B_1=1)- P(A_1=B_1=1).
\end{align}
See the \hyperlink{methods}{Methods} section for formal definitions and
proofs.

This inequality avoids the detector-efficiency loophole so that
experimental violation does not need the fair-sampling assumption.
However, it does not take into account how pairs are identified,
e.g., how coincidences are determined by timing data in the
experimental output. We would want to establish a similar inequality
that includes restricting to a subset of pairs,
$P(A_j=B_k=1\cap\Lambda_{A_jB_k})$, and detection, e.g.,
$P(A_j=1\cap\Lambda_{A_j})$. Fortunately, this is not too difficult.

There are two alternative methods. The first uses fixed
\emph{non-overlapping} time slots,
\begin{equation}
  \label{eq:slots}
  S_i=\big\{t:t_i\le t < t_i+\tau\big\},
\end{equation}
for detection and coincidence determination (we use the same $\tau$
for time slot size and time window size because that gives a similar
rate of accidental coincidences, making the two methods easily
comparable).  This enables a CH-type inequality that avoids the
coincidence-time loophole, making experiments that use the new
inequality independent of the fair-coincidence assumption.  The reason
is that a fixed time slot border treats long and short delays equally,
see \hyperlink{fig:2}{Fig.~\ref{fig:2}b}. In pulsed photonic
experiments, there is a natural time slot structure because of the
pulse timing. However, also for continuously pumped experiments fixed
time slots for coincidence identification can be easily enforced.

A detection is only counted if it occurs in one of the time slots, in
a local realist model corresponding to
\begin{equation}
  \begin{split}
    \Lambda_{A_j}(i) =\big\{\lambda:T_{A}(a_j,\lambda)\in S_i\big\},\\
    \Lambda_{B_k}(i) =\big\{\lambda:T_{B}(b_k,\lambda)\in S_i\big\},
  \end{split}
\end{equation}
and for all the time slots we have the disjoint union
\begin{equation}
  \Lambda_{A_j}
  =\bigcup_i\Lambda_{A_j}(i),
\end{equation}
and similar for $\Lambda_{B_k}$. A coincidence occurs in slot $i$ if
both detections occur there, and this happens on the set
$\Lambda_{A_j}(i)\cap\Lambda_{B_k}(i)$. A coincidence in any slot
occurs on the disjoint union
\begin{equation}
  \Lambda_{A_jB_k}
  =\bigcup_i\Big(\Lambda_{A_j}(i)\cap\Lambda_{B_k}(i)\Big).
\end{equation}
This is not the factor structure we have in the detector
efficiency case, but it will enable us to recover the appropriate
inequality. It is important that time slots are assigned locally, so
that no remote influence is present; such influence would result in
a set structure similar to that in Eq.~\eqref{eq:6}. We can now
include coincidence determination in the inequality, and arrive at
\begin{align}
 \label{eq:CH}
    P&(A_1=B_2=1\;\cap\;\Lambda_{A_1B_2})%\\&\quad
    + P(A_2=B_1=1\;\cap\;\Lambda_{A_2B_1})\notag\\
    &\quad-P(A_2=B_2=1\;\cap\;\Lambda_{A_2B_2})\notag\\
    &\le P(A_1=1\;\cap\;\Lambda_{A_1})+P(B_1=1\;\cap\;\Lambda_{B_1})\notag\\
    &\quad-P(A_1=B_1=1\;\cap\;\Lambda_{A_1B_1}).
\end{align}

The key observation here is that the inequality avoids the
coincidence-time loophole and can be properly violated by experiment,
as soon as disjoint fixed time slots are used. It does not matter how
the time slots are chosen (as long as they are locally assigned), or
if they have a natural counterpart in the experiment. This is
especially important in continuously pumped photonic experiments.

We should point out that the event $A_j=1$ does not necessarily mean a
single click in a detector in one time slot. It is a label for some
event we are interested in. In the data analysis, one can choose the
event $A_j=1$ to correspond to \emph{at least one} detection for
setting $a_j$ at the first site, and similarly for $B_k=1$ (setting
$b_k$ at the second). Note that this choice must be made entirely from
the locally available information. A coincidence (the event
$\Lambda_{A_jB_k}$) would then correspond to (any number of)
detections in two equal-indexed time slots, using information from
both sites. Operationally, this means to coarse-grain all detector
clicks on each side and for every time slot to dichotomic values:
``0'' = ``no detections,'' ``1'' = ``one or more detections.'' This
coarse-graining method allows no switching of settings within time
slots, but this can be handled also.

The alternative method does not need fixed time slots. The intuition
for this method is that all proposed local realist models that
exploit the loophole put some of the $A_2B_2$ detection events more
than $\tfrac{1}{2} \tau$ apart, so that they are not identified as a
coincidence. It therefore makes sense to increase the window for the
$A_2B_2$ events. With
\begin{equation}
  \label{eq:triple}
  \begin{split}
    \Lambda_{A_1B_1} &= \big\{\lambda:\big|T_{A}(a_1,\lambda)
    -T_{B}(b_1,\lambda)\big|<\tfrac{1}{2} \tau_1\big\},\\
    \Lambda_{A_1B_2} &= \big\{\lambda:\big|T_{A}(a_1,\lambda)
    -T_{B}(b_2,\lambda)\big|<\tfrac{1}{2} \tau_2\big\},\\
    \Lambda_{A_2B_1} &= \big\{\lambda:\big|T_{A}(a_2,\lambda)
    -T_{B}(b_1,\lambda)\big|<\tfrac{1}{2} \tau_3\big\},\\
    \Lambda_{A_2B_2} &= \big\{\lambda:\big|T_{A}(a_2,\lambda)
    -T_{B}(b_2,\lambda)\big|<\tfrac{1}{2} \textstyle\sum_i \tau_i\big\},\\
  \end{split}
\end{equation}
we obtain
\begin{equation}
  \Lambda_{A_1B_1}\cap\Lambda_{A_1B_2}\cap\Lambda_{A_2B_1}\;\subset\;\Lambda_{A_2B_2}.
\end{equation}
In other words: if the $A_1B_1$, $A_1B_2$, and $A_2B_1$ detection events
that are separated by at most $\tau_i$ are identified as coincidences,
then the $A_2B_2$ detection events separated by at most $\sum_i \tau_i$
are also identified as coincidences (see
\hyperlink{fig:2}{Fig.~\ref{fig:2}c}). This gives the inequality
\begin{align}
  \label{eq:CHsum}
    P&(A_1=B_2=1\;\cap\;\Lambda_{A_1B_2}) %\\&\quad
    + P(A_2=B_1=1\;\cap\;\Lambda_{A_2B_1})\notag\\
    &\quad-P(A_2=B_2=1\;\cap\;\Lambda_{A_2B_2})\notag\\
    &\le P(A_1=1)+P(B_1=1)\notag\\&\quad-P(A_1=B_1=1\;\cap\;\Lambda_{A_1B_1}).
\end{align}

We remark that, in contrast to the ``fixed-time-slots method''
(\hyperlink{fig:2}{Fig.~\ref{fig:2}b}), no coarse-graining of multiple
clicks is applied in the ``window-sum method''
(\hyperlink{fig:2}{Fig.~\ref{fig:2}c}).

\begin{figure}[t]
  \centering
  \hypertarget{fig:3}{}%
  \includegraphics[width=.98\linewidth]{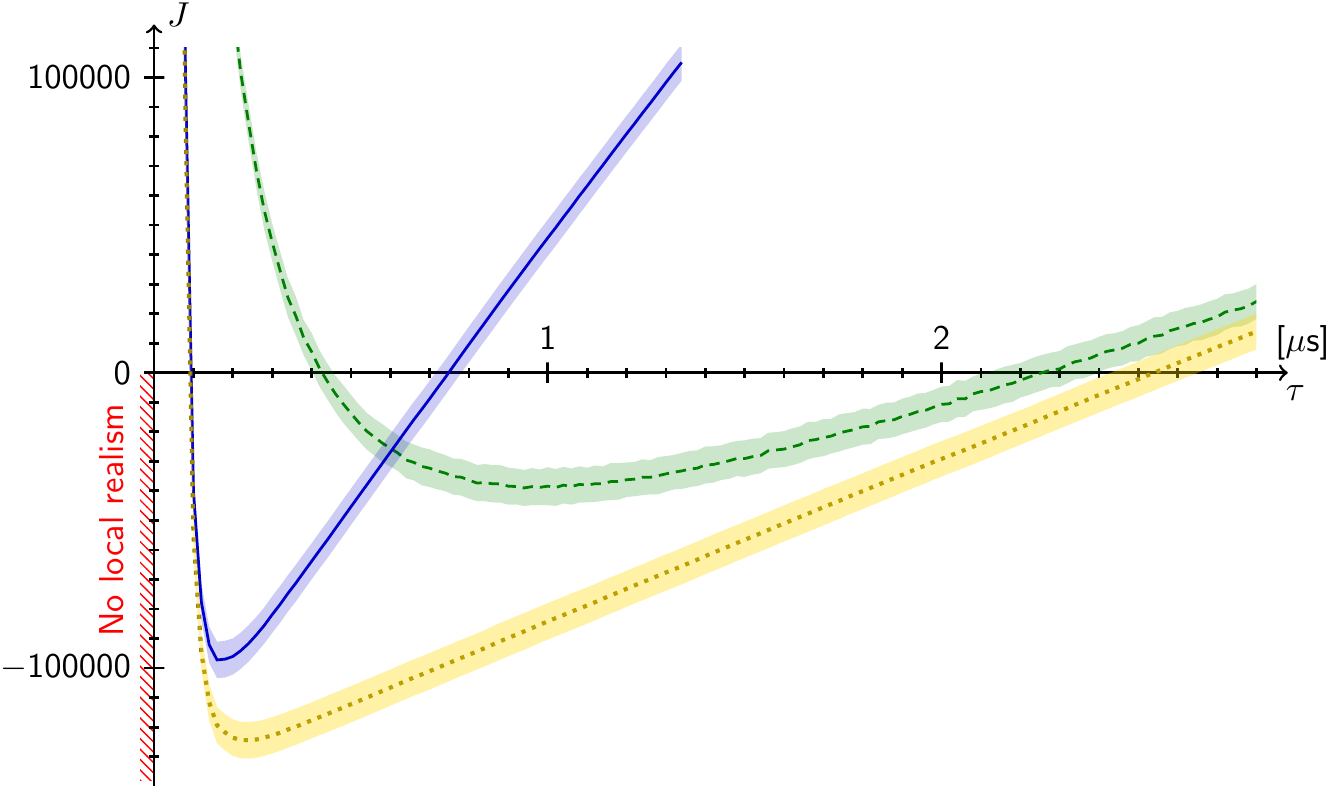}
  \caption{Experimental $J$ values plotted as a function of
    coincidence window or time slot size $\tau$. The three
    alternatives are \textbf{a)}~the dotted yellow line that uses the
    moving-windows method and the fair-coincidence assumption;
    \textbf{b)}~the green dashed line that uses the fixed-time-slots
    method; and \textbf{c)}~the blue continuous line that uses the
    window-sum method (with all $\tau_i$ equal to $\tau$). The latter
    two are not vulnerable to the coincidence-time loophole. The
    shading corresponds to plus/minus three estimated standard
    deviations. Negative $J$ values cannot be explained by local
    realist models.}
  \label{fig:3}
\end{figure}

\textbf{Violation from the experiment \cite{Gius2013} ---} The CH
inequality is not vulnerable to the detector-efficiency loophole,
and therefore free of the fair-sampling assumption. As shown above,
similar inequalities that are not vulnerable to the coincidence-time
loophole can be derived by two methods (Ineq.~\eqref{eq:CH} and
\eqref{eq:CHsum}), both therefore free of the fair-coincidence
assumption. Both statements also hold for the Eberhard inequality,
where all probabilities in the CH inequality are replaced by the
corresponding number of counts. Applying the fixed-time-slots or
window-sum method leads to Eberhard-type inequalities similar to
Ineq.~\eqref{eq:CH} and \eqref{eq:CHsum}, free of the
fair-coincidence assumption as well as the fair-sampling assumption.
Collecting all terms on one side creates inequalities of the form
\begin{equation}
  \label{eq:4}
  J \ge 0.
\end{equation}

Since the fixed-time-slots method removes coincidences as compared
to the moving windows method (compare
\hyperlink{fig:2}{Fig.~\ref{fig:2}a} and
\hyperlink{fig:2}{Fig.~\ref{fig:2}b}), it comes at a cost in a
continuously pumped experiment: because of the inherently random
emission times and the timing jitter of the detectors, two photons
close in time that would be coincident in the moving-window method
may belong to different time slots and fail to register a
coincidence in the fixed-time-slots method.  To minimize the loss,
long adjacent slots (much longer than the timing jitter of the
detectors used) are desirable, and the earlier-mentioned
coarse-graining should be used because multiple generated pairs can
appear in the same slot. Coarse-grained coincidences may not show
quantum correlations, and can prohibit or hamper a violation of the
tested Bell inequality. Therefore, depending on experimental
parameters such as timing jitter, overall efficiency, background
counts, and rate of generated pairs, there will be an optimal size
for the locally predefined time slots, see
\hyperlink{fig:2}{Fig.~\ref{fig:3}b}.

For the experiment \cite{Gius2013}, choosing adjacent predefined
time slots with size $\tau = 980$~ns, yields $J= -38803 \pm
2020$. This corresponds to a violation, free of the fair-coincidence
assumption, by more than $19\,\sigma$ (estimated standard
deviations). The standard deviation is estimated analogously to
\cite{Gius2013}, by dividing the data set into 30 subsets and using
standard unbiased point estimates, adjusting these to apply to the
sum of the 30 samples rather than the mean.

Finally, we also analyze the data of Ref.~\cite{Gius2013} using the
window-sum method (\hyperlink{fig:2}{Fig.~\ref{fig:2}c} and
\hyperlink{fig:2}{Fig.~\ref{fig:3}c}) with all three $\tau_i$ being
equal to $\tau$ and the window for $A_2B_2$ being $3 \tau$. For
$\tau = 180$~ns, one obtains $J = -96988 \pm 2076$, a $46\,\sigma$
violation (estimated standard deviations). The window-sum method
typically leads to a larger violation than the fixed-time-slots
method since it evades the trade-offs encountered in choosing a slot
size. The only ``penalty'' is an increase of the accidental
coincidences for the $A_2B_2$ events. Therefore, the window-sum
method can be a valuable tool in situations where unfavorable
experimental parameters (such as high timing jitter and dark counts
of the detectors) do not allow a violation using the
fixed-time-slots method.

For the above $J$ values we calculated the singles counts as
follows: Alice's singles counts for outcome $A_1$ were taken to be
the mean of her singles when she and Bob applied settings $a_1,b_1$
and when they applied $a_1,b_2$. Similarly, Bob's singles counts for
outcome $B_1$ were taken to be the average obtained from the setting
combinations $a_1,b_1$ and $a_2,b_1$. In an ideal experiment, the
$J$ value should not depend on whether this averaging is employed or
individual combinations are taken. However, due to the residual
drifts discussed in Ref.~\cite{Kofl2013} the above reported $J$
values for the data of Ref.~\cite{Gius2013} do depend on the
procedure, albeit not in any way that would alter the conclusion,
namely a significant violation of CH- or Eberhard-type inequalities
that are not vulnerable to the coincidence-time loophole.

\textbf{Conclusion ---} In their original form, the CH and Eberhard
inequalities are derived without using the assumption of fair
sampling. Here, we have derived CH- and Eberhard-type inequalities
that are also free of the coincidence-time loophole, through two
different approaches. One is to use fixed time slots for the local
measurement results and identify coincidences if detections occur in
equal time slots, while the other is to choose a key coincidence
window as long as the sum of the others. Both methods can be used in
continuously pumped experiments as well as in pulsed experiments, and
in particular, both can be used to show that the experiment reported
in Ref.~\cite{Gius2013} violates local realism and is not vulnerable
to the coincidence-time loophole, therefore not needing the
fair-coincidence assumption.

\textbf{\hypertarget{methods}{Methods} ---} Realist models (or hidden
variable models) are probabilistic models that use three building
blocks. The first building block is a sample space $\Lambda$, which is
the set of possible hidden variable values. The second is a family of
event subsets $E\subset\Lambda$, e.g., the sets of hidden variable
values where specific measurement outcomes occur. The third and final
building block is that these event sets must be measurable using a
probability measure $P$, so that each event has a well-defined
probability. If the sample (hidden variable) $\lambda$ is reasonably
well-behaved, its distribution $\rho$ can be constructed from
this. Below, we will use deterministic realism without loss of
generality, since stochastic realist models are equivalent to mixtures
of deterministic ones \cite{Fine1982}. We can now prove the following
theorems.

\smallskip\noindent \emph{Theorem 1 (Clauser-Horne):} The following
three prerequisites are assumed to hold except at a null set:
\begin{enumerate}[i]
\item \label{C1} \emph{Realism.} Measurement results can be described
  by two families of random variables ($A$ for Alice with local
  setting $a$, $B$ for Bob with local setting $b$):
  \begin{equation}
        A(a,b,\lambda)\text{ and }B(a,b,\lambda).
  \end{equation}
  The dependence on the hidden variable $\lambda$ is usually
  suppressed in the notation.
\item \label{C2} \emph{Locality.}  Measurement results are independent
  of the remote setting:
  \begin{equation}
    \begin{split}
      A(a,\lambda)&\stackrel{\textrm{def}}= A(a,b_1,\lambda)= A(a,b_2,\lambda)\\
      B(b,\lambda)&\stackrel{\textrm{def}}= B(a_1,b,\lambda)= B(a_2,b,\lambda).
    \end{split}
  \end{equation}
  For brevity we define $A_j(\lambda)=A(a_j,\lambda)$ and
  $B_k(\lambda)=B(b_k,\lambda)$.
\item \label{C3} \emph{Freedom of choice.} The measurement setting
  distribution does not depend on the hidden variable, or
  equivalently, the probability measure $P$ does not depend on the
  measurement settings; this is sometimes formulated as independence
  between the distribution $\rho$ of $\lambda$ and the measurement
  settings,
  \begin{equation}
    P(E|a,b)=P(E)\text{ or }\rho(\lambda|a,b)=\rho(\lambda).
  \end{equation}
\end{enumerate}
Then,
\begin{align}
    P&(A_1=B_2=1) + P(A_2=B_1=1)- P(A_2=B_2=1)\notag\\
    &\le P(A_1=1) + P(B_1=1)- P(A_1=B_1=1).
\end{align}

\smallskip\noindent \emph{Proof:}
\begin{align}
    P&(A_1=B_2=1) + P(A_2=B_1=1) - P(A_2=B_2=1)\notag\\
    &\le P(A_1=B_2=1) + P(A_2=B_1=1)\notag\\
    &\quad -P(A_1=B_2=1\cap A_2=B_1=1)\notag\\
    &= P(A_1=B_2=1\;\cup\;A_2=B_1=1)\notag\\
    &\le P(A_1=1\;\cup\;B_1=1)\\
    &= P(A_1=1) + P(B_1=1)- P(A_1=B_1=1)\tag*{$\square$}
\end{align}

Using the notation for fixed {non-overlapping} time slots from the
main text, we have the following theorem.

\smallskip\noindent \emph{Theorem 2 (The Clauser-Horne inequality with
  disjoint time slots):} If the prerequisites from the Clauser-Horne
inequality are assumed to hold except at a null set, and also:
\begin{enumerate}[iv]
\item \label{Cc4} \emph{Disjoint time slots.}  Detections are obtained
  on subsets $\Lambda_{A_j}$; $\Lambda_{B_k}$ with $j,k = 1,2$ of $\Lambda$ that are disjoint unions of the form
  \begin{equation}
    \Lambda_{A_j}=\bigcup_i\Lambda_{A_j}(i),
  \end{equation}
  and coincidences are obtained on subsets $\Lambda_{A_jB_k}$ of $\Lambda$ that
  are disjoint unions of the form
  \begin{equation}
    \Lambda_{A_jB_k}=\bigcup_i\Big(\Lambda_{A_j}(i)\cap\Lambda_{B_k}(i)\Big).
  \end{equation}
\end{enumerate}
Then
\begin{align}
    P&(A_1=B_2=1\;\cap\;\Lambda_{A_1B_2}) + P(A_2=B_1=1\;\cap\;\Lambda_{A_2B_1})\notag\\
    &\quad-P(A_2=B_2=1\;\cap\;\Lambda_{A_2B_2})\notag\\
    &\le P(A_1=1\;\cap\;\Lambda_{A_1})+P(B_1=1\;\cap\;\Lambda_{B_1})\notag\\
    &\quad-P(A_1=B_1=1\;\cap\;\Lambda_{A_1B_1}).
\end{align}

\noindent \emph{Proof:} Define new time-slot-indexed random
variables, e.g., $A_{j}^{(i)}$ that indicates that $A_j=1$ in time
slot $i$,
\begin{equation}
  \label{eq:1}
  A_{j}^{(i)}=1 \quad\Leftrightarrow\quad\big(A_j=1\cap\Lambda_{A_j}(i)\big).
\end{equation}
Since the sets are disjoint unions, we have for example
\begin{equation}
  \begin{split}
    P(A_1=1\;\cap\;\Lambda_{A_1})
    &=P\Big(A_1=1\;\cap\;\big(\bigcup_i\Lambda_{A_1}(i)\big)\Big)\\
    &=\sum_iP(A_{1}^{(i)}=1),
  \end{split}
\end{equation}
and
\begin{equation}
  \begin{split}
    P&(A_1=B_2=1\;\cap\;\Lambda_{A_1B_2})\\
    &=P\Big(A_1=B_2=1\;\cap\;\big(\bigcup_i\Lambda_{A_1}(i)\cap\Lambda_{B_2}(i)\big)\Big)\\
    &=\sum_iP\big(A_{1}^{(i)}=B_{2}^{(i)}=1\big).
  \end{split}
\end{equation}
The inequality now follows from using the original CH inequality in
each time slot. \hfill$\square$

Finally, for the window-sum method we have the following.

\smallskip\noindent \emph{Theorem 3 (The Clauser-Horne inequality with a
  subset property):} If the prerequisites from the Clauser-Horne
inequality are assumed to hold except at a null set, and also:
\begin{enumerate}[iv]
\item \label{Cs4} \emph{Subset property.}  Coincidences are obtained
  on subsets $\Lambda_{A_1B_1}$; $\Lambda_{A_1B_2}$; $\Lambda_{A_2B_1}$; and
  $\Lambda_{A_2B_2}$, of $\Lambda$, and the last coincidence set contains
  the intersection of the other three,
\begin{equation}
  \label{eq:subset}
  \Lambda_{A_1B_1}\cap\Lambda_{A_1B_2}\cap\Lambda_{A_2B_1}\;\subset\;\Lambda_{A_2B_2}.
\end{equation}
\end{enumerate}
Then
\begin{align}
  P&(A_1=B_2=1\;\cap\;\Lambda_{A_1B_2}) + P(A_2=B_1=1\;\cap\;\Lambda_{A_2B_1})\notag\\
  &\quad-P(A_2=B_2=1\;\cap\;\Lambda_{A_2B_2})\notag\\
  &\le P(A_1=1)+P(B_1=1)\notag\\
  &\quad-P(A_1=B_1=1\;\cap\;\Lambda_{A_1B_1}).
\end{align}

\noindent \emph{Proof:} We need to treat the $A_2B_2$ events
separately,
\begin{equation}
  \label{eq:3}
  \begin{split}
    &P(A_2=B_2=1\;\cap\;\Lambda_{A_2B_2})\\
    &\;\ge P(A_2=B_2=1\;\cap\;\Lambda_{A_1B_1}\cap
    \Lambda_{A_1B_2}\cap\Lambda_{A_2B_1})\\
    &\;\ge
    P\Big(\big((A_1=B_2=1\cap\Lambda_{A_1B_2})\\
    &\qquad\quad\cap (A_2=B_1=1\cap\Lambda_{A_2B_1})\big) \cap\Lambda_{A_1B_1}\Big).
  \end{split}
\end{equation}
The proof of Theorem 1 now applies on the subset $\Lambda_{A_1B_1}$
so that
\begin{equation}
  \begin{split}
    P&(A_1=B_2=1\;\cap\;\Lambda_{A_1B_2}\cap\Lambda_{A_1B_1})\\
    &\quad + P(A_2=B_1=1\;\cap\;\Lambda_{A_2B_1}\cap\Lambda_{A_1B_1})\\
    &\quad-P(A_2=B_2=1\;\cap\;\Lambda_{A_2B_2})\\
    &\le P(A_1=1\cap\Lambda_{A_1B_1})
    + P(V_1=1\cap\Lambda_{A_1B_1})
    \\&\quad -P(A_1=B_1=1\;\cap\;\Lambda_{A_1B_1}).
  \end{split}
\end{equation}
Since
\begin{equation}
  \begin{split}
    P&(A_1\!=\!B_2\!=\!1\cap\Lambda_{A_1B_2}\setminus\Lambda_{A_1B_1})
    \le P(A_1\!=\!1\setminus\Lambda_{A_1B_1}),\\
    P&(A_2\!=\!B_1\!=\!1\cap\Lambda_{A_2B_1}\setminus\Lambda_{A_1B_1})
    \le P(B_1\!=\!1\setminus\Lambda_{A_1B_1}),
  \end{split}
\end{equation}
the result follows.
\hfill$\square$

\emph{Note added ---} During the preparation of this manuscript it has
come to our attention that the window-sum method has been discovered
independently by Emanuel Knill and co-workers in the context of an
extended statistical analyis~\cite{Knil2013}.

\emph{Acknowledgments ---} We would like to acknowledge discussions
with Anton Zeilinger, Sae Woo Nam, and Emanuel Knill; and assistance
with the detection system from Alexandra Mech, Jörn Beyer, Adriana
Lita, Brice Calkins, Thomas Gerrits, and Sae Woo Nam. SR is supported
by an EU Marie-Curie Fellowship (PIOF-GA-2012-329851).

\end{document}